\documentclass[conference]{IEEEtran}
\IEEEoverridecommandlockouts
\usepackage{cite}
\usepackage{amsmath,amssymb,amsfonts}
\usepackage{amsthm}
\usepackage{subcaption}
\usepackage{graphicx}
\usepackage{textcomp}
\usepackage{xcolor}
\usepackage{xspace}
\usepackage{booktabs}

\usepackage{optidef}
\usepackage{algorithm}
\usepackage{algpseudocode}
\usepackage{lipsum}
\def\BibTeX{{\rm B\kern-.05em{\sc i\kern-.025em b}\kern-.08em
    T\kern-.1667em\lower.7ex\hbox{E}\kern-.125emX}}

\begin{document}
\makeatletter
\newcommand*{\algrule}[1][\algorithmicindent]{%
  \makebox[#1][l]{%
    \hspace*{.2em}
    \vrule height .75\baselineskip depth .25\baselineskip
  }
}

\newcount\ALG@printindent@tempcnta
\def\ALG@printindent{%
    \ifnum \theALG@nested>0
    \ifx\ALG@text\ALG@x@notext
    \else
    \unskip
    \ALG@printindent@tempcnta=1
    \loop
    \algrule[\csname ALG@ind@\the\ALG@printindent@tempcnta\endcsname]%
    \advance \ALG@printindent@tempcnta 1
    \ifnum \ALG@printindent@tempcnta<\numexpr\theALG@nested+1\relax
    \repeat
    \fi
    \fi
}
\patchcmd{\ALG@doentity}{\noindent\hskip\ALG@tlm}{\ALG@printindent}{}{\errmessage{failed to patch}}
\patchcmd{\ALG@doentity}{\item[]\nointerlineskip}{}{}{} 
\makeatother

\title{A Lightweight Machine Learning Approach for Delay-Aware Cell-Switching in 6G HAPS Networks}

\author{\IEEEauthorblockN{Görkem Berkay Koç$^{1,2}$, Berk Çiloğlu$^{1,2}$, Metin Ozturk$^{1,2}$, Halim Yanikomeroglu$^2$}
\IEEEauthorblockA{$^1$Electrical and Electronics Engineering, Ankara Yıldırım Beyazıt University, Ankara, Türkiye\\
\IEEEauthorblockA{$^2$Non-Terrestrial Networks (NTN) Lab, Systems and Computer Engineering, Carleton University, Ottawa, Canada}}
}
\maketitle

\begin{abstract}
This study investigates the integration of a high altitude platform station (HAPS), a non-terrestrial network (NTN) node, into the cell-switching paradigm for energy saving. By doing so, the sustainability and ubiquitous connectivity targets can be achieved.
Besides, a delay-aware approach is also adopted, where the delay profiles of users are respected in such a way that we attempt to meet the latency requirements of users with a best-effort strategy. 
To this end, a novel, simple, and lightweight $Q$-learning algorithm is designed to address the cell-switching optimization problem.
During the simulation campaigns, different interference scenarios and delay situations between base stations are examined in terms of energy consumption and quality-of-service (QoS), and the results confirm the efficacy of the proposed $Q$-learning algorithm.
\end{abstract}


\section{Introduction}
One of the usage scenarios under the wing of the sixth generation of communication systems (6G) is to ensure ubiquitous connectivity~\cite{6G_ubiquitious}, which can be an achievable target with the dense deployment of terrestrial networks (TNs); however this, in turn, conflicts with the sustainability promises---another key aspect that 6G is dedicated to enhance---since it increases the energy consumption of cellular networks. 
In this regard, non-terrestrial networks (NTNs) are capable of addressing both ubiquitous connectivity and sustainability challenges, and therefore 6G aims at integrating NTN and TN~\cite{6G_ubiquitious}. 
High-altitude platform station (HAPS) is one of the upcoming member of NTNs operating as a super macro base station (SMBS)\footnote{International Telecommunication Union (ITU) refers HAPS-SMBS as high-altitude International Mobile Telecommunications (IMT) base station (HIBS)~\cite{itu_vision_june_23}.} with its unique position in the stratosphere---being closer to the ground than satellites~\cite{haps_smbs}. 

After the integration, cell-switching (CS) methods, which aim to put the base stations (BSs) which are not in use (or lightly used) into a sleep mode and offload their loads~\cite{profit, ELAA2022JR, metin}, can be applied in order to reduce the energy consumption of TNs.  
In order to deactivate a BS, at least two conditions need to be satisfied: 
There must be another BS in a close proximity (e.g., neighbor BS) to be used for offloading purposes, and that neighbor BS needs to have available resource blocks to accommodate the offloaded load. 
At this point, HAPS-SMBS can be an operable remedy with its gargantuan footprint, which is between 40 km and 100 km~\cite{main_survey}, and massive capacity. 
Therefore, the integration of TN and NTN paves the way for meeting the expectations of ubiquitous connectivity and better sustainability towards 6G networks, as these two items are included in the 6G framework of ITU~\cite{itu_vision_june_23}.

Another usage scenario of 6G is \textit{hyper-reliable and low-latency communication (HRLLC)}, an extension to the prominent \textit{ultra-reliable and low-latency communication (URLLC)} scenario of fifth generation of communication systems (5G)~\cite{itu_vision_june_23}. 
To this end, it can be foreseen that networks have to incur less delay (e.g., mission-critical scenarios) while satisfying ubiquitous connectivity and sustainability commitments.
In other words, with 6G, network optimization becomes even more complicated task as there are multiple conflicting objectives in the equation along with several stringent constraints. 
In this work, we address these ubiquitous connectivity, sustainability, and delay challenges all together by incorporation a HAPS-SMBS in the CS workaround and making the system delay-aware by respecting the latency requirements of users with the best-effort strategy.
Therefore, this work is an important attempt in realizing the promises of 6G networks. 

Although there is an abundant amount of research on CS in TN in the literature, HAPS-supported CS in cellular networks is not mature yet. 
An earlier version of this paper in~\cite{bizim} proposed a CS approach for HAPS-integrated TN using the exhaustive search (ES) technique.
Sorting of small BS (SBS) loads in ascending order was studied in~\cite{meryem} for offloading traffic to macro BS (MBS) and HAPS-SMBS, while the authors in~\cite{yeni_haps} examined a scenario where all the traffic was offloaded to the HAPS-SMBS.
Moreover, in \cite{cihan}, the authors studied HAPS-SMBS to accommodate the unexpected traffic demands of users in TN by analyzing the capital and operational expenditures, while how the presence of HAPS-SMBS could affect sustainability in the environment of traditional micro and macro BSs was presented in~\cite{canhaps6g}.

In this paper, we build a CS optimization model considering a HAPS-SMBS in the network and then develop a $Q$-learning algorithm to solve this optimization problem.
The designed $Q$-learning algorithm is lightweight and easy to implement, because one of the objectives of this work is to analyze whether a very simple algorithm can solve this complex optimization problem (proven to be NP-hard~\cite{nphard_proven}).
We acknowledge that there have been abundant of advanced algorithms proposed in the literature that solve various CS problems~\cite{profit, ELAA2022JR, metin}, the idea behind proposing such a lightweight algorithm is to find a good trade-off between accuracy and computational complexity as well as ease of implementation.
While considering the optimization of energy efficiency, we also consider delay-sensitive UEs (DSUs) in our model to satisfy their requirements of quality-of-service (QoS). 
Additionally, this paper is one of the very first attempts in the literature that considers a delay-aware approach while investigating the effect of various interference scenarios under the scope of the NTN-integrated CS concept. 

The contributions of this study are summarized as follows:
\begin{itemize}
    \item We build a CS optimization model for a network where HAPS-SMBS and MBS co-exist to minimize the energy consumption of the network.
    \item We propose a novel delay-aware CS approach for NTN-assisted systems to investigate the impact of the delay profiles of the user equipments (UEs).
    \item We design a lightweight $Q$-learning algorithm, named Reduced Action-State Pair (RASP), to find a reasonable solution for the developed optimization problem.
    \item Different network models that vary according to the carrier frequency of network elements (e.g., SBS, MBS, and HAPS-SMBS) are also examined in order to investigate the effects of interference on the CS performance.
\end{itemize}


\section{System Model and Problem Formulation}
\subsection{Network Model}
\begin{figure}
   \centering
    \includegraphics[width=.8\linewidth]{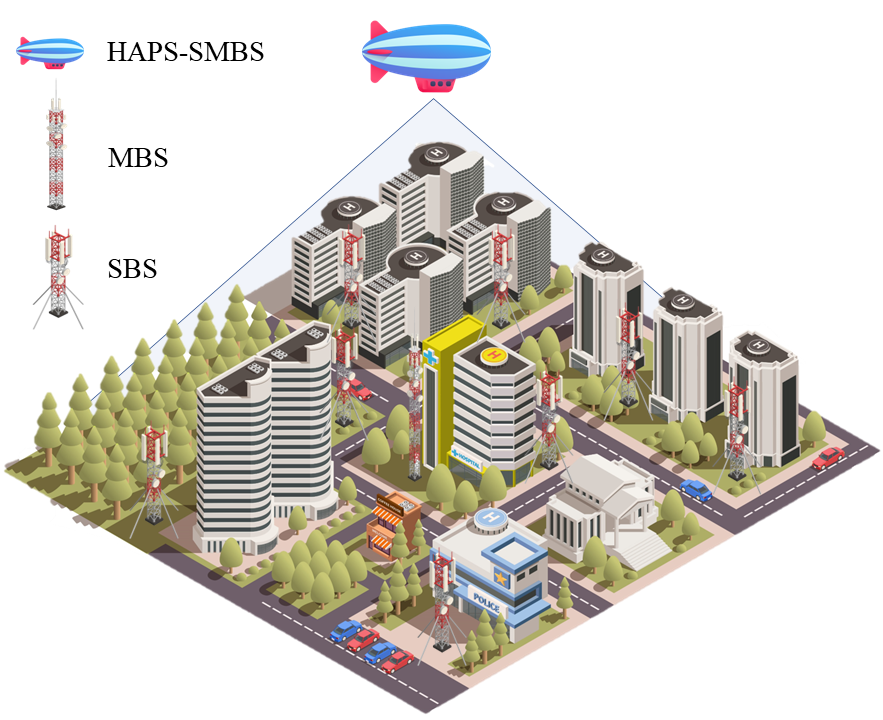}
    \caption{A VHetNet containing SBSs, MBS, and HAPS-SMBS.}
    \label{fig:networkmodel}
\end{figure}
A vertical heterogeneous network~(VHetNet) architecture consisting $n \in \mathbb{N}$ SBSs, with $k=\{1, 2, ..., n\}$ keeps the indices (i.e., IDs) of SBSs, an MBS, and a HAPS-SMBS are used as a network model.
There are also $e \in \mathbb{N}$ users/UEs, with $i=\{1, 2, ..., e\}$ holding the indices for the users/UEs, distributed uniformly in the considered environment.
MBS is located at the center of the environment, while SBSs are symmetrically positioned with respect to MBS, and HAPS is placed 20 km above the ground by centralizing the considered environment.
Note that we assume that 70\% of the HAPS-SMBS capacity is reserved for the considered network, whereas the rest of 30\% is used by other networks within the coverage area of HAPS-SMBS.
The system model is illustrated in Fig.~\ref{fig:networkmodel}, wherein four UE mobility types are considered: stationary, pedestrian, cyclist, and driver.


\subsection{Propagation Model}
The path loss model of HAPS-SMBS is obtained according to the 3rd Generation Partnership Project (3GPP) report given in~\cite{3GPP_HAPS}.
By taking the line-of-sight (LoS) and non-LoS (NLoS) scenarios into the account, the path loss $L^\omega$, $\omega \in \text{\{LoS,~NLoS\}}$, is obtained as~\cite{3GPP_HAPS}
\begin{equation}\label{eq:PL2}
    L^\omega = L^\omega_\text{b} + L_\text{g} + L_\text{s} + L_\text{e},
\end{equation}
where $L^\omega_\text{b}$ is the basic path loss, $L_\text{g}$ is the attenuation due to the atmospheric gasses, $L_\text{s}$ is the loss caused by either ionospheric or tropospheric scintillation, and $L_\text{e}$ denotes building entry loss. 
More details about the path loss model of HAPS-SMBS is given in~\cite{bizim}.
The 3GPP report given in~\cite{3GPP_Terrestrial} is considered for TN for the coherency between NTN and TN path loss models.

\subsection{User and Base Station Delay Models}
In this work, each user is associated with a delay profile according to the application that they run.
In particular, a binary classification is adopted, where the users are referred to as delay sensitive~(i.e., DSU) and non-DSU~(NDSU).
As such, let $H^\text{u}_{i}\in \{0,1\}$ denote the delay sensitivity of users, such that $H^\text{u}_{i}=1$ if UE $i$ is DSU, and $H^\text{u}_{i}=0$ otherwise. 
Besides, the BSs included in the network are also classified according to their delay profiles (i.e., the end-to-end delay caused by using a particular BS, $H^\text{b}_{j}\in \{0,1\}$) in a way that the TN BSs (i.e., SBS and MBS) are considered to incur lower end-to-end delay(i.e., $H^\text{b}_{j}=0$) than the NTN BS (i.e., $H^\text{b}_{j}=1$).
Therefore, it is preferable to associate DSUs to TN BSs, while NDSUs can be assigned to both TN and NTN BSs.

In addition, a dissatisfaction variable, $\theta_i \in \{0,1\}$, indicating the satisfaction status of user $i$, such that $\theta_i=0$ if the user is satisfied and $\theta_i=1$ if the user is dissatisfied, which only happens if a DSU is associated with HAPS-SMBS.
Then, the total dissatisfaction count, $\theta_\text{T}$, is computed as $\theta_\text{T} = \sum^e_{i=1}\theta_i$.


\subsection{User-Cell Association}\label{sec:user_assoc}
Even though users' signal-to-interference-plus-noise ratio (SINR) levels are of a prime importance for user-cell association process, there are two first level criteria~(FLC), and two second-level criteria~(SLC) to associate a user $i$ to BS $j$:
\begin{itemize}
    \item \textbf{[FLC-1]} \textbf{Capacity availability}: $\{\varkappa_1\in \{0,1\}=1~|~\Omega^\text{a}_j \geq \lambda_i\}$, where $\Omega^\text{a}_j, \lambda_i\in \mathbb{R^+}$ denote the available capacity of BS $j$ and user demand of UE $i$ at time $t$, respectively, while $\Omega_j\in \mathbb{R^+}$ represents the total capacity of BS $j$.
    \item \textbf{[FLC-2]} \textbf{Receiver sensitivity}: $\{\varkappa_2\in \{0,1\}=1~|~P_{\text{RX},i} \geq P_{\text{RX}_\text{min},i}\}$, where $P_{\text{RX},i}$ and $P_{\text{RX}_\text{min},i}$ indicate the received power and receiver sensitivity, respectively.
    \item \textbf{[SLC-1]} \textbf{Delay profile}: A user is preferably associated to a BS that matches its delay profile: i.e., a DSU is associated to a TN BS. $\{\varkappa_3\in \{0,1\}=1~|~H^\text{b}_{j}=H^\text{u}_{i}\}$
    \item \textbf{[SLC-2]} \textbf{Provision of the maximum SINR}: $\{\varkappa_4\in \{0,1\}=1~|~\gamma_{j} > \forall \gamma_{h\neq j},~1 \leq j,h \in \mathbb{N}\leq n+2\}$, where $\gamma_j$ represents the instantaneous SINR of BS $j$. 
\end{itemize}

Regarding the criteria, FLC are the one that are obligatory be met, while SLC are preferable.
More specifically, a BS needs to satisfy FLCs, followed by attempting to satisfy SLC with a \textit{best-effort} strategy, such that the UE selects the BS that yields the maximum SINR (SLC-2) among the ones that satisfy SLC-1.
The detailed user allocation process is given in Algorithm~\ref{algo:user_alloc}.
Let $U_{i,j}\in \{0,1\}$ represent the association between UE $i$ and BS $j$, such that $U_{i,j}=1$ if UE $i$ can be associated to BS $j$, and $U_{i,j}=0$ otherwise.

\begin{algorithm} 
\caption{Delay-Aware User Allocation Algorithm} \label{algo:user_alloc}
\begin{algorithmic}[1]
\scriptsize

\For{$i=1:e$ (each user)} 
    \For{$j=1:n+2$ (each BS)}
        \If{$H^\text{u}_{i}=1$}
             \If{$\Omega_j \geq \lambda_i$ ($j \neq \text{HAPS-SMBS}$) and $P_{\text{RX},i} \geq P_{\text{RX}_\text{min},i}$} 
                \State Obtain $~\gamma_{j}$ where $~\gamma_{j} > \forall \gamma_{h\neq j}$ 
            \Else
                \Comment{No capacity in the terrestrial BSs.}
                \If{$\Omega_j \geq \lambda_i$ ($j = \text{HAPS-SMBS}$) and $P_{\text{RX},i} \geq P_{\text{RX}_\text{min},i}$}
                    \State $U_{i,j}=1$
                    \Comment{Allocate to HAPS-SMBS.}
                    \State $\theta_i=1$ 
                    \Comment{The user is dissatisfied.}
                \Else
                    \Comment{No capacity in all BSs.}
                    \State $U_{i,j}=0$
                    \Comment{The user is out of service.}
                \EndIf
            \EndIf    
        \Else 
            \Comment{UE is non-DSU}
            \If{$\Omega_j \geq \lambda_i$ and $P_{\text{RX},i} \geq P_{\text{RX}_\text{min},i}$}
                \State Obtain $~\gamma_{j}$ where $~\gamma_{j} > \forall \gamma_{h\neq j}$
            \Else
                \State $U_{i,j}=0$
                \Comment{The user is out of service.}
            \EndIf
        \EndIf    
    \EndFor
\EndFor

\end{algorithmic}
\end{algorithm}

\subsection{Power Consumption Model}
The EARTH model in~\cite{earth} is adopted as the power consumption model to calculate $P_\text{B}^\alpha$ where $\alpha \in \text{\{S, M,~H\}}$.
Hereafter, any variable with superscript S, M, or H indicates that the variable is given for SBSs, MBS, and HAPS-SMBS, respectively.
The power consumption of a BS is formulated as
\begin{equation}\label{eq:pow1}
P_\text{B}^\alpha = 
    \begin{cases}
        P_\text{C}^\alpha + \xi^\alpha \rho^\alpha P_\text{max}^\alpha, & 0 < P_\text{TX}^\alpha < P_\text{max}^\alpha, \\
        P_\text{S}^\alpha, & P^\alpha_\text{TX}= 0,
    \end{cases}
\end{equation}
where $P_\text{C}^\alpha$ is constant power, $\xi^\alpha$ slope of the load-dependent power consumption value, $\rho^\alpha=[0,1]$ is the load of BSs.
$P_\text{TX}$ and $P_\text{max}$ are the instantaneous and the maximum transmit powers of BSs, respectively.
$P_\text{S}^\alpha$ is the power consumption when a BS is in the sleep mode. 

\subsection{Problem Formulation}\label{sec:problem}

This study aims to find the best policy that yields minimum energy consumption of the network. A policy $\pmb \eta_t = [\beta_1, \beta_2, \beta_3, ...,\beta_n]$ expresses which SBSs are active/deactive at time $t$. $\beta_{k,t} \in \{0,1\}$ represents active/deactive states of SBS $k$ at time $t$. It should be noted that HAPS-SMBS and MBS are supposed to be always ON.

This combinatorial model can be formulated as~\cite{bizim}
\begin{mini}|s|
    {\eta}{P_\text{N}} 
    {\label{eq:opt}}{}
    \addConstraint{\rho^\alpha}{\leq1}{\qquad\text{(C$_1$)}}
    \addConstraint{P_\text{TX}^\alpha}{\leq P_\text{max}^\alpha}{\qquad\text{(C$_2$)}}
    \addConstraint{H^\text{b}_{j}}{=H^\text{u}_{i},~~\forall i,j}{\qquad\text{(C$_3$)},}
\end{mini} 
where C$_1$ ensures that the BSs (SBS, MBS, and HAPS-SMBS) accommodate loads under their maximum capacity while C$_2$ is responsible for keeping the BSs transmitting at a power below their maximum allowance.
Moreover, C$_3$ states that there are UEs need to be satisfied with their delay requirements.

The total power consumption of the VHetNet is given as
\begin{equation}\label{eq:totalpower}
P_\text{N} = P^\text{H}_\text{B} + P^\text{M}_\text{B} + \sum_{k=1}^{n}P^\text{S}_{\text{B},k},
\end{equation}
where $P^\text{S}_{\text{B},k}$ shows the power consumption of the $k^\text{th}$ SBS.
Substituting \eqref{eq:pow1} into \eqref{eq:totalpower} and doing $P^\text{T}_\text{B}=P^\text{H}_\text{B} + P^\text{M}_\text{B}$ results

\begin{equation}\label{eq:closed-form-power}
    P_\text{N} =P^\text{T}_\text{B}+\sum_{k=1}^{n} (P^\text{S}_{\text{C},k} + \xi^\text{S}_k \rho^\text{S}_k P^\text{S}_{\text{max},k})\beta_{k,t} + P^\text{S}_{\text{S},k}(1-\beta_{k,t}).
\end{equation}

\section{Methodology}
Besides the ES algorithm, which tries every possible combination of CS and finds the one resulting in the minimum energy consumption, the use of artificial intelligence (AI)-based algorithms in 6G wireless communication will increase as stated in the ITU-R framework in~\cite{itu_vision_june_23} that AI will play a key role in shaping 6G networks.
In this regard, the use of reinforcement learning (RL) is considered in this study; more specifically $Q$-learning, which is an RL algorithm based on the consequences of actions, is employed. 
In $Q$-learning, each action has an associated penalty or reward, and the results of the taken actions are stored in a matrix called $Q$-table, which consists of all the actions (as rows/columns) and all states (as columns/rows); hence when the number of states and/or columns becomes huge, it would be hard (if not impossible) to store such a large matrix, with which performing mathematical operations is another challenge to overcome.

In order to solve this problem, one option is to use deep RL algorithms.
Nonetheless, due to the complicated nature of deep learning, we propose utilizing a simple and lightweight $Q$-learning algorithm, which is more advantageous in terms of computational cost. 
Therefore, the scope of this paper is not developing a very accurate algorithm for the CS problem, as there is an avalanche of studies in the literature about that. The objective of this paper is rather investigating whether a simple and lightweight algorithm is able to solve this complex (i.e., NP-hard) optimization problem, because the cellular communications networks have continuously becoming more complex and there is an urgent need for simpler solutions.

\subsection{Reduced Action-State Pair (RASP) Design} 
In this design, the states, $\pmb S$, and actions, $\pmb A$, that make up the $Q$-table are intended to be reduced so that the $Q$-table can be simplified. 
The main idea behind this design is to reduce both $\pmb S$ and  $\pmb A$ at the same time in order to minimize the computational complexity.

The action design is inspired by the fact that any natural number can be written in terms of the summation of some powers of two. 
This fact lies on a similar ground with Fourier's theorem, which, roughly speaking, states that any periodic signal can be written in terms of sinusoidal functions.
This concept can be formulated mathematically as $x = \sum_{n=-\infty}^{\infty} \text{a}_n 2^n$, assuming that $x$ is a real number and a$_n \in \{0,1\}$ is a coefficient.
Using the the active/deactive status of SBSs given by $\pmb \eta_t = [\beta_1, \beta_2, \beta_3, ...,\beta_n]$, a binary number is created as $\beta_{b} = \beta_1\beta_2...\beta_n$, where each digit corresponds to the active/deactive status of each SBS.
Inspired by~\cite{metin}, within the proposed method, the set of possible
actions is defined as: $\pmb A = \{0, \pm2^0, \pm2^1,..., \pm2^n\}$.
Thus, taking an action at time $t$ refers $\beta_{d,t+1} =\beta_{d,t} + A_x$ where $A_x$ comes from $\pmb A$ and $\beta_d$ is decimal form of $\beta_b$. 
The set of states, $\pmb S$, is based on the number of BSs exceeding the whole capacity; such that $s=\sum_{j=1}^{n+2}\Lambda_j$, where $\Lambda_j\in \{0,1\}$ is 1 if BS $j$ exceeds the full capacity and 0 otherwise.
The computational complexity of this design is $O(n^2)$. 

As a penalty function for RASP design is $\Gamma =  \zeta \times P_\text{N} + (1-\zeta)\times \Phi$, where $\zeta = \{\zeta_1,\zeta_2,...,\zeta_n\}$ is load of all SBSs, and $\zeta=0$ if $\sum_i \sum_j U_{i,j}<e$ and $\zeta=1$ if $\sum_i \sum_j U_{i,j}=e$, where $\Phi$ is an astronomical constant penalty to prevent selecting an action that leads to a disconnected UE.

    
            

\section{Performance Evaluation}

To implement the delay-aware NTN-integrated CS simulations, MATLAB is used as the development environment where we run three different scenarios, namely Scenarios A, B, and C. 
Scenario A is a configuration in which SBSs and HAPS-SMBS are on the same carrier frequency (i.e., interference limited) while the MBS is on a different carrier frequency (i.e., noise-limited).
This scenario represents a favorable MBS link---as it is not affected by the interference---in a delay-aware network since DUSs demand to connect a terrestrial BS. 
In Scenario B, all types of BSs (SBS, MBS, and HAPS-SMBS) are on the same carrier frequency. 
The main purpose of including this scenario in the analysis is to observe how the CS algorithms (proposed and benchmark) behave in a network where all BSs interfere with each other.
Scenario C is the same with Scenario B in terms of the carrier frequencies, but the transmit power and antenna gain of HAPS-SMBS are set to the those used for MBS. 
In other words, in Scenario C, HAPS-SMBS is disadvantaged in order to discover the impacts of transmit power and antenna gain of HAPS-SMBS.
Besides the ES algorithm, all-active approach (A3) is used as a simulation benchmark where whole SBSs are always kept active.
Table~\ref{table:simparameter} includes both communication and $ Q$-learning-related simulation parameters.

\begin{table}[hb]
\centering
\caption{Simulation Parameters} \label{table:simparameter}
\resizebox{.9\columnwidth}{!}{%
\begin{tabular}{ll}
\textbf{Parameters} & \textbf{Values} \\ \hline
Environment area    & 1000 m $\times$ 1000 m               \\
Time slot number ($N_\text{TS}$) and duration ($t_\text{d}$)  & 25 \& 1 second                    \\
Interference-limited carrier frequency (HAPS-SMBS, SBS, MBS)   & 2.5 GHz                   \\
Noise-limited carrier frequency (MBS - Scenario A)    & 3 GHz         \\
Bandwidth  ($W$) and bandwidth per UE  & 50 MHz \& 200 kHz          \\
Transmit power: SBSs \& MBS    & 33 dBm \& 46 dBm       \\
Transmit power: HAPS-SMBS---Scenario A, B \& Scenario C   & 49 dBm \cite{3GPP_Terrestrial} \& 46 dBm            \\
Antenna gain: SBSs \& MBS    & 4 dBi \& 17 dBi         \\
Antenna gain HAPS-SMBS---Scenario A, B \& Scenario C  & 43.2 dBi \cite{3GPP_HAPS}   \& 17 dBi      \\
Antenna gain: UE     & 0 dBi  \cite{3GPP_Terrestrial}          \\
$\sigma^\text{LoS}$ \& $\sigma^\text{NLoS}$ &  4 dB \& 6 dB \cite{3GPP_Terrestrial}  \\
Receiver reference sensitivity  &  -95 dBm  \\
Constant power: SBSs ($P_\text{C}^\text{SBS}$)             & 56 W           \\
Constant power: MBS ($P_\text{C}^\text{M}$) \& HAPS-SMBS ($P_\text{C}^\text{H}$)      & 130 W           \\
Slope of load-dependent power: SBSs ($\xi^\text{SBS}$)             & 2.6       \\ 
Slope of load-dependent power: MBS ($\xi^\text{M}$) \& HAPS-SMBS ($\xi^\text{H}$)        & 4.7       \\ 
Maximum transmit power: SBSs ($P_\text{max}^\text{SBS}$)    & 6.3 W \\
Maximum transmit power: MBS ($P_\text{max}^\text{M}$) \& HAPS-SMBS ($P_\text{max}^\text{H}$)   & 20 W \\
Sleep mode power ($P_\text{s}$)    & 39 W \\
Learning rate    & 0.9  \\
Discount factor  & 0.9 \\
Initial exploration and exploitation criteria ($\epsilon$) & 0.8 \\
Iteration number & 2000 \\
\bottomrule
\end{tabular}
}
\end{table}

\subsection{Performance Metrics}

Three different metrics are utilized to evaluate the performance, which are energy consumption, relative difference, and data rate.
\textit{Energy consumption} values are obtained by multiplying the power consumption values, computed through~\eqref{eq:pow1}, with the time slot duration ($t_\text{d}$).
\textit{Relative difference} is defined as the percentage difference between the energy consumption values of the algorithms. 
The lower value is subtracted from the greater one (always written first), followed by a division by the greater one.
\textit{Data rate} presents the performance of the CS algorithms in terms of the quality of communication. The values are obtained using Shannon's capacity formula at the user side. 
It is an important metric because CS tends to lower the user data rates.

\subsection{Results and Discussions}

The network is simulated for three different user densities, i.e., users$/\text{m}^2$ ratio (denoted by $\delta$) values, in order to investigate both sparse and dense network conditions. 
The value of $\delta$ is varied by three $e$ values (i.e., the number of UEs), such that $e=600$, $e=1200$, and $e=2400$.  
Moreover, two delay profile distributions are considered as $D = 30\%$ and $D = 70\%$, that is, 30\% and 70\% of the UEs are DSUs, respectively.


\begin{figure*}
    \begin{subfigure}[t]{.3\textwidth}
    \includegraphics[width=\linewidth]{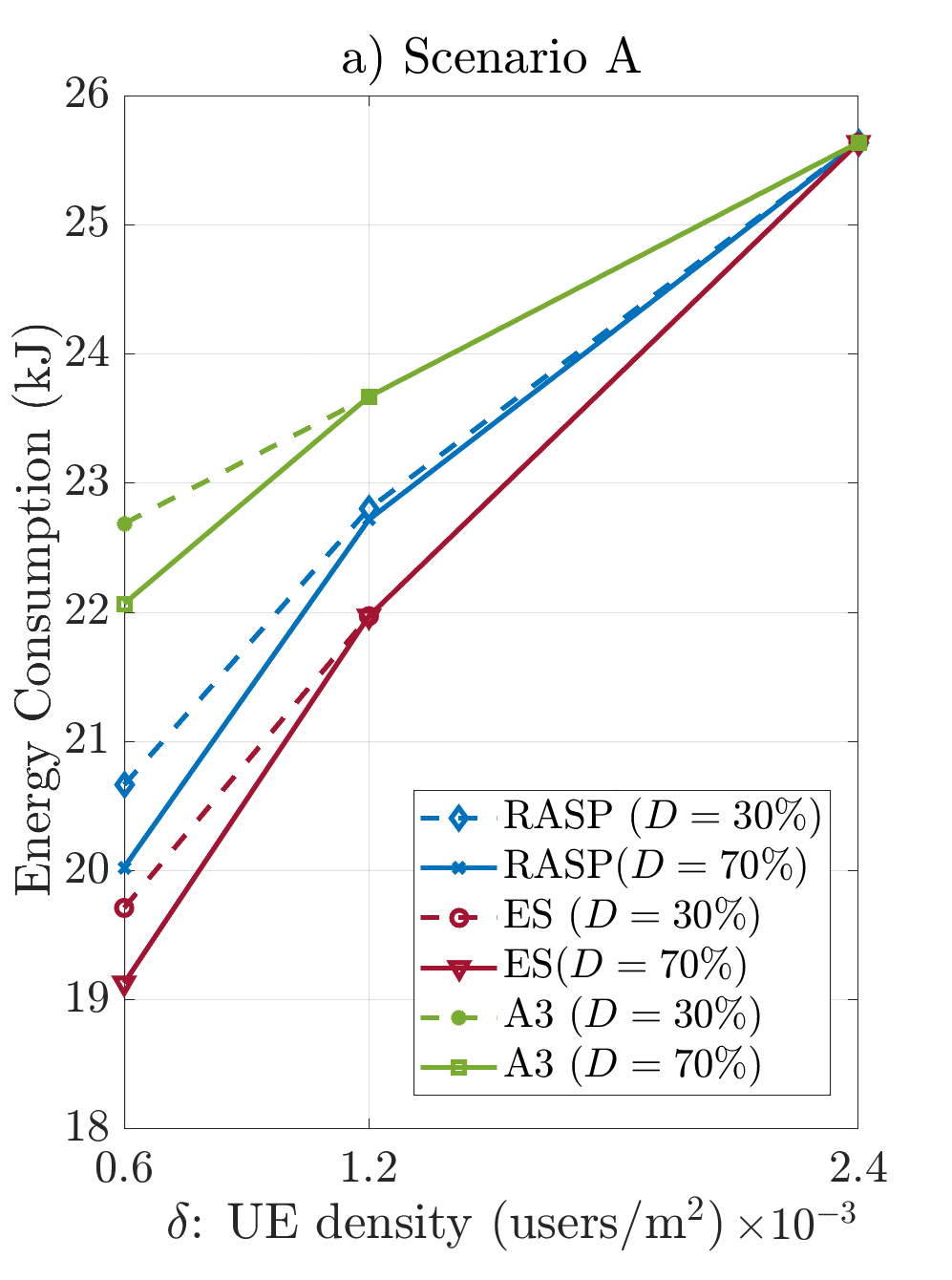}
    \label{Energy_A}
    \end{subfigure}
    \qquad
   \begin{subfigure}[t]{.3\textwidth}
    \includegraphics[width=\linewidth]{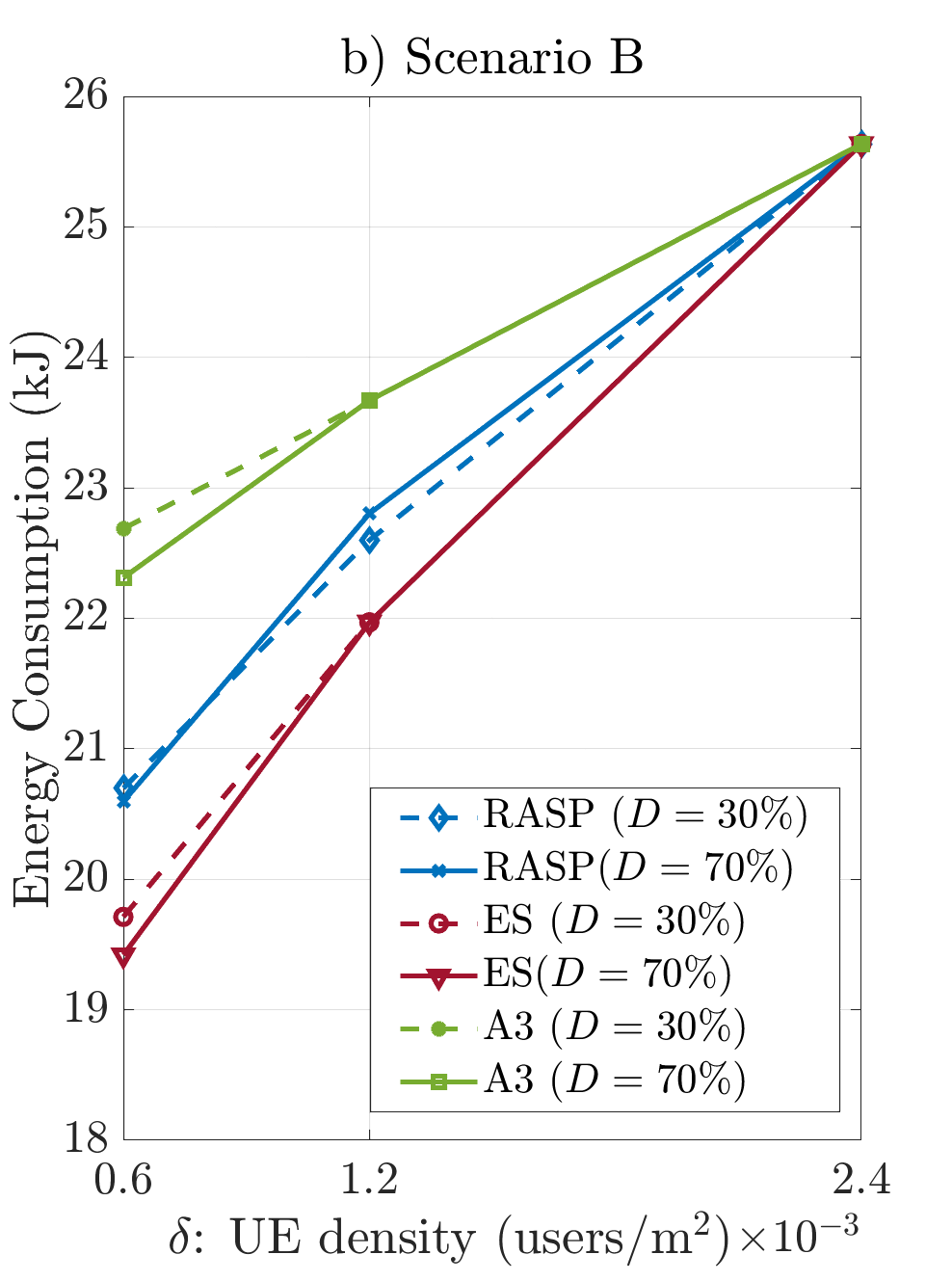}
    \label{Energy_B}
    \end{subfigure}
    \qquad
    \begin{subfigure}[t]{.3\textwidth}
    \includegraphics[width=\linewidth]{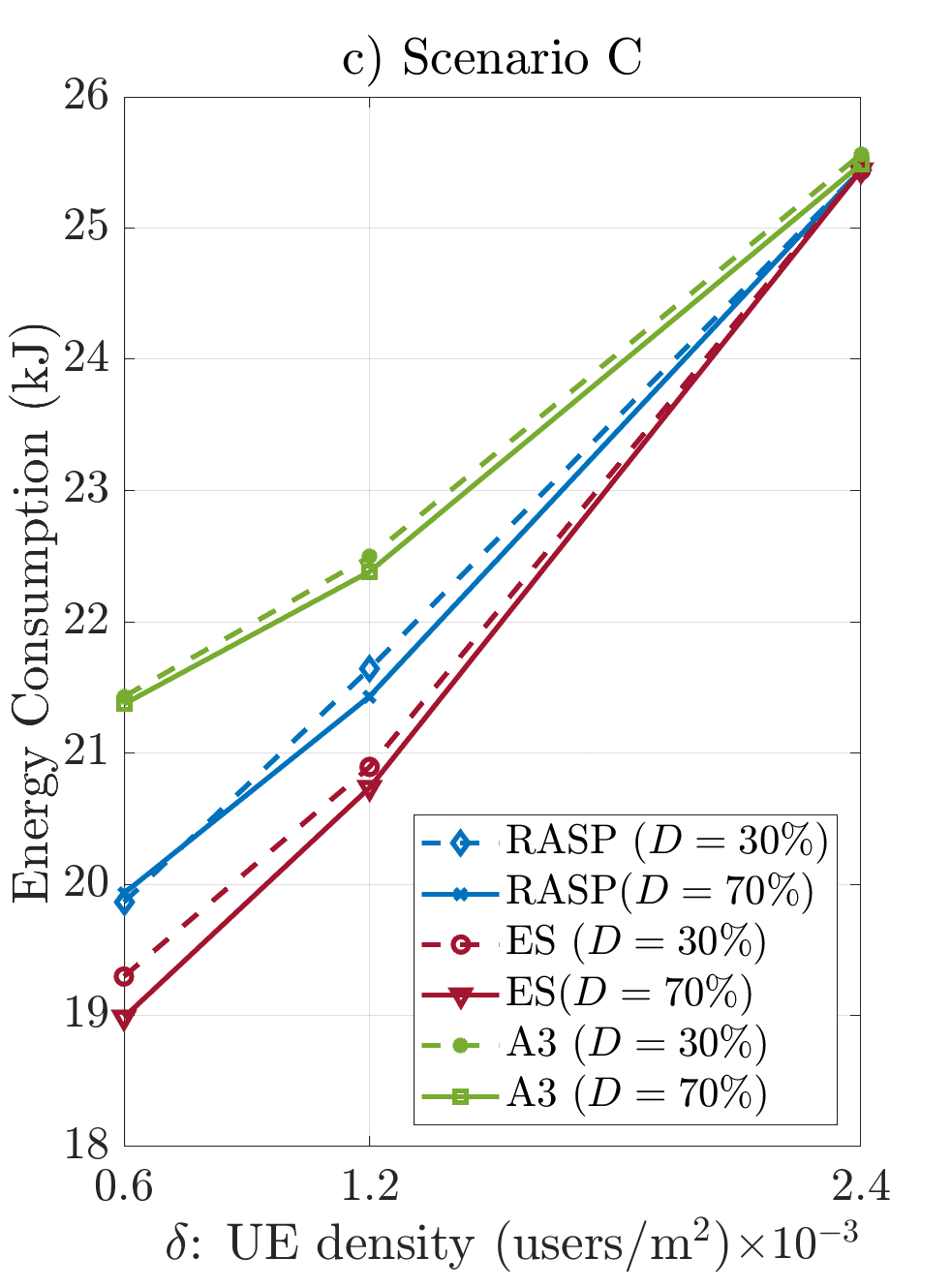}
    \label{Energy_C}
    \end{subfigure}
\caption{Energy consumption of different algorithms and comparison for different scenarios.}
    \label{energy}
\end{figure*}

Fig. \ref{energy} shows the energy consumption of the network for different $\delta$ values and all scenarios.
The most fundamental takeaway is that the difference between A3 and other two algorithms (i.e., ES and RASP) is wide for the lower user densities, and the difference narrows as the user density rises. 
This is quite an expected result since as the network becomes dense, there exist less switching-off opportunities and more BSs become active, leading to increased energy consumption.
Specifically, when $\delta=0.6\times 10^{-3}$, the relative difference between A3 and other algorithms is at its maximum of 13.3\% among all scenarios and delay profiles ($D = 30\%$ and $D = 70\%$), whereas it drops to almost 0\% when $\delta=2.4\times 10^{-3}$.

Considering all the scenarios, it can be seen that for Scenarios A and B, both delay profile distributions give the same energy consumption result when $\delta \ge 1.2\times10^{-3}$. 
The rationale behind this is that HAPS-SMBS and MBS present better signal links for these scenarios, and they are the top two favorable allocation stations for the users, thereby users connect to SBSs only after both HAPS-SMBS and MBS become fully loaded.
It should be noted that as HAPS-SMBS and MBS have higher load-dependent factor values ($\xi^\text{H}$ and $\xi^\text{M}$, respectively) than that of SBSs ($\xi^\text{SBS}$), allocating a user to HAPS-SMBS or MBS causes more energy consumption.
The reason why the results for $D = 30\%$ and $D = 70\%$ are not exactly the same when $\delta=0.6\times10^{-3}$ is that MBS can accommodate all DSUs as its capacity is adequate to host all DSUs when $D = 30\%$, and offers a satisfactory signal quality due to its transmit power and antenna gain properties.
However, it does not have enough resource blocks to host all DSUs when $D = 70\%$; there can be user allocation fluctuations between $D = 30\%$ and $D = 70\%$ at $\delta=0.6\times10^{-3}$.
Revisiting the results of $\delta=1.2\times10^{-3}$, since MBS is not sufficient, we do not expect a difference between $D = 30\%$ and $D = 70\%$ as HAPS-SMBS and MBS provide strong links to the users in Scenarios A and B.

Unlike Scenarios A and B, the energy consumption results are different between $D = 30\%$ and $D = 70\%$ for $\delta=1.2\times10^{-3}$ in Scenario C. 
Since Scenario C represents the outcomes when the HAPS-SMBS is disadvantaged, HAPS-SMBS is no more capable of providing a favorable link to the ground users---also it is located 20 km above from the ground (i.e., a very high path loss). This also affects the interference occurrences between scenarios. At this point, SBSs become more favorable than HAPS-SMBS for most users, and therefore the energy consumption when $D = 30\%$ is higher since it has fewer DSUs, and the rest of the users are offloaded to HAPS-SMBS, subsequently a slight difference occurs when Scenario C is compared to the other two scenarios.

To evaluate the RASP algorithm design, it can be observed that even though the results do not converge to the ES algorithm, the obtained results are reasonable. 
RASP makes up to 36\% the same switching decision with the ES algorithm in Scenario B for $D = 30\%$ and $\delta=0.6\times10^{-3}$.
It is observable that the relative difference between RASP and ES drops as user density rises from $\delta=0.6\times10^{-3}$ to $\delta=1.2\times10^{-3}$ for both Scenarios A and B, when $D = 30\%$ is considered. The least drop takes place in Scenario A with a relative difference of 0.96\% whereas the largest drop occurs in Scenario B with a relative difference of 1.98\%. As an opposite result, there is a 0.6\% relative difference increment between RASP and ES for $\delta=0.6\times10^{-3}$ and $\delta=1.2\times10^{-3}$ within the context of Scenario C.
When $D = 70\%$, the relative difference between RASP and ES decreases for all three scenarios as the user density increases from $\delta=0.6\times10^{-3}$ to $\delta=1.2\times10^{-3}$, meaning that the RASP algorithm exhibits a closer performance to the ES algorithm when $D = 70\%$; Scenario A has the least decrease with 1.18\% while Scenario B has the biggest decrease with 2.06\% relative difference.
 


Analyzing the A3 algorithm for $D = 30\%$ and $D = 70\%$ leads us to observe that the highest relative difference is obtained as high as 2.73\% when $\delta=0.6\times10^{-3}$ for Scenario A while the lowest is 0.27\% for Scenario C at the same user density. The reason for this is that HAPS-SMBS is disadvantaged in Scenario C. 
As SBSs offer favorable links to users due to the reduced interference from their perspective, NDSUs may prefer connecting to SBSs instead of HAPS-SMBS.
Regarding the ES algorithm, the highest relative difference between $D = 30\%$ and $D = 70\%$ is found as 2.96\% when Scenario A is in place for $\delta=0.6\times10^{-3}$, whereas the least relative difference becomes the 1.48\% at the same user density. 
The reason why the behavior of the ES algorithm between $D = 30\%$ and $D = 70\%$ differs from the behavior of A3 between $D = 30\%$ and $D = 70\%$ is that ES switches off some SBSs according to the given user density while A3 algorithm keeps all SBSs active regardless of the user density, making A3 more prone to be affected by interference.


\begin{figure}
    \begin{subfigure}{.5\textwidth}
    \includegraphics[width=.91\linewidth]{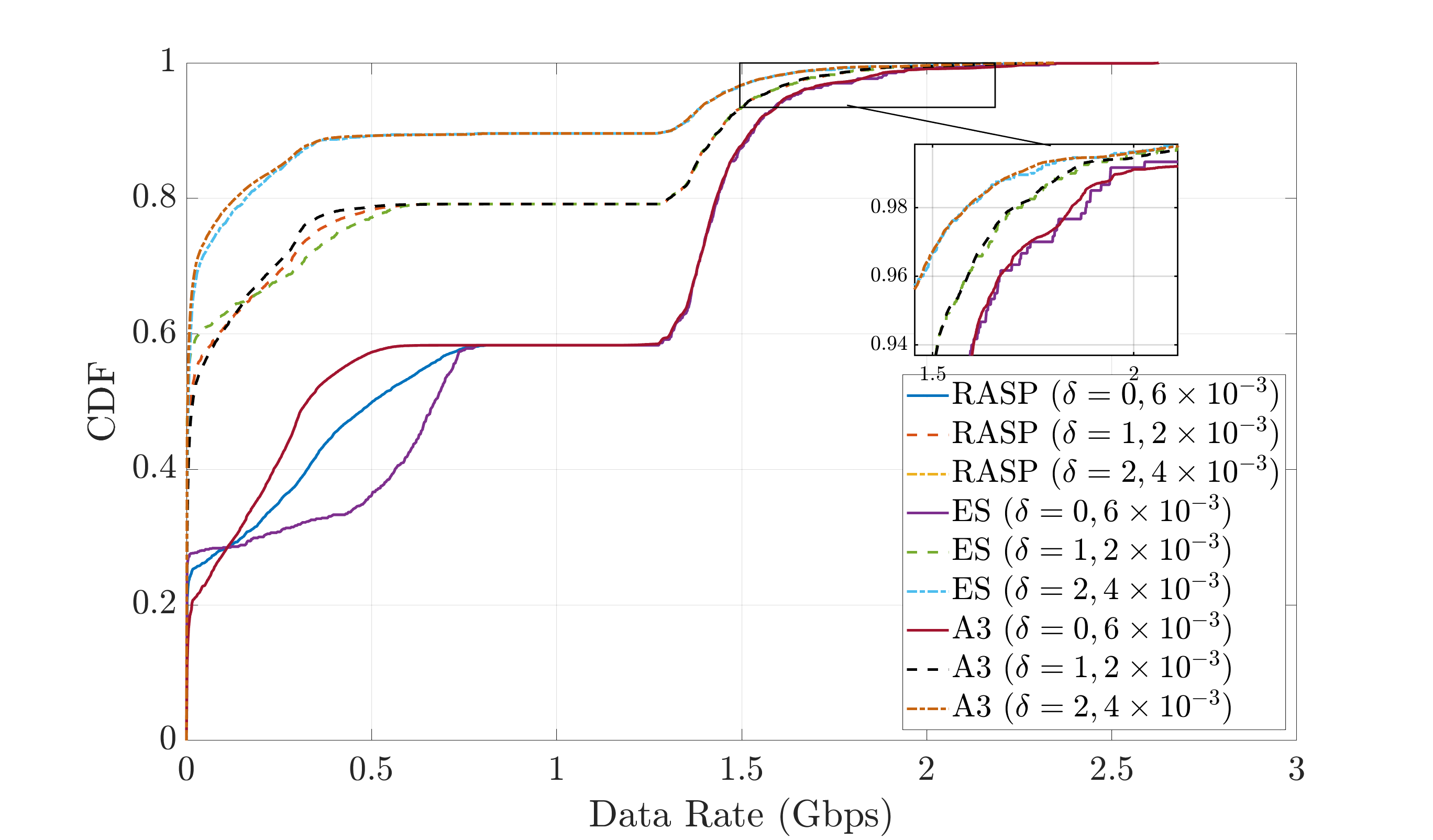}
    \caption{Scenario A when $D = 30\%$.}
    \label{rate_A}
    \end{subfigure}
    
  \begin{subfigure}{.5\textwidth}
    \includegraphics[width=.91\linewidth]{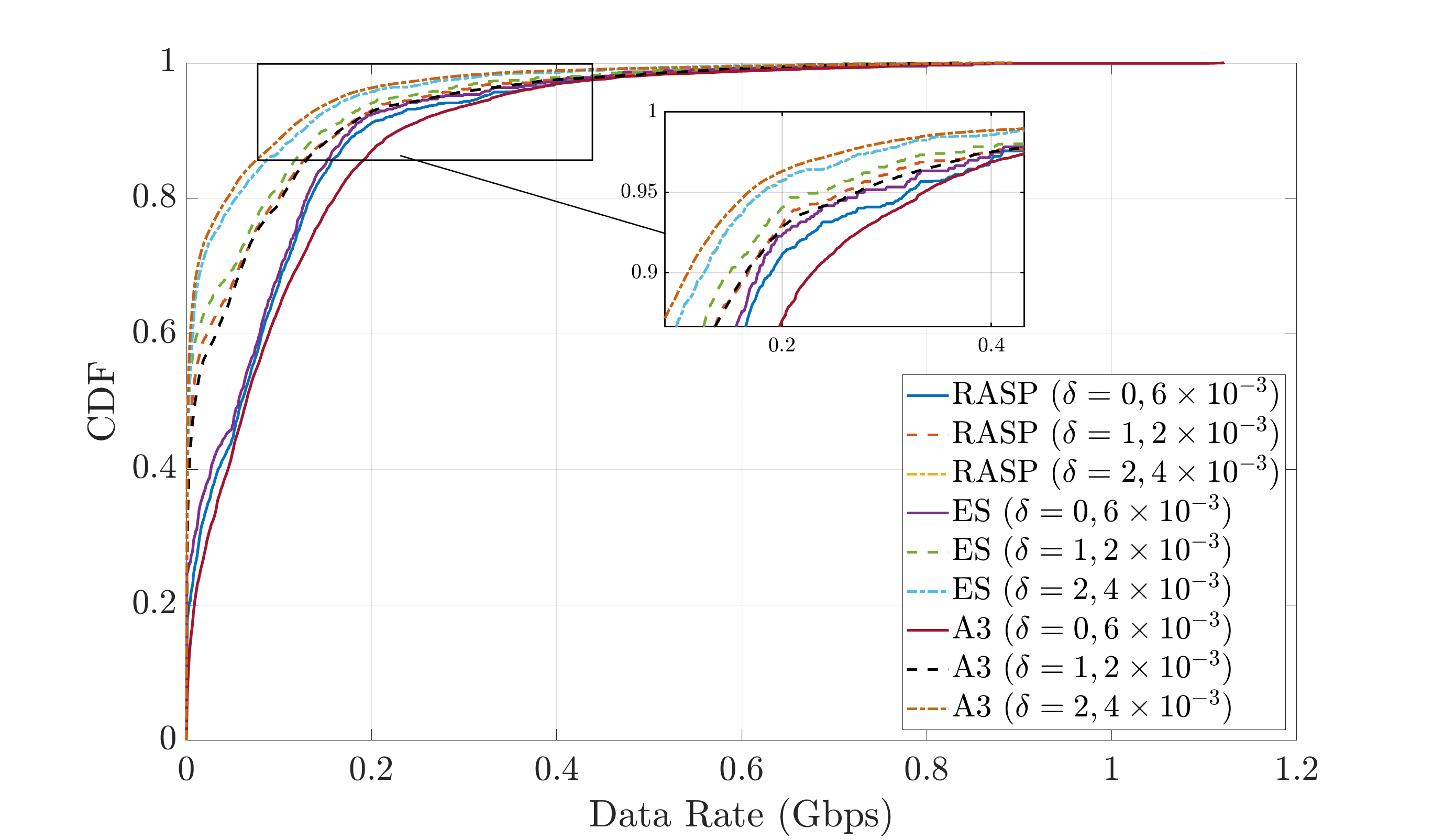}
    \caption{Scenario B when $D = 30\%$.}
    \label{rate_B}
    \end{subfigure}
    
     \begin{subfigure}{.5\textwidth}
    \includegraphics[width=.91\linewidth]{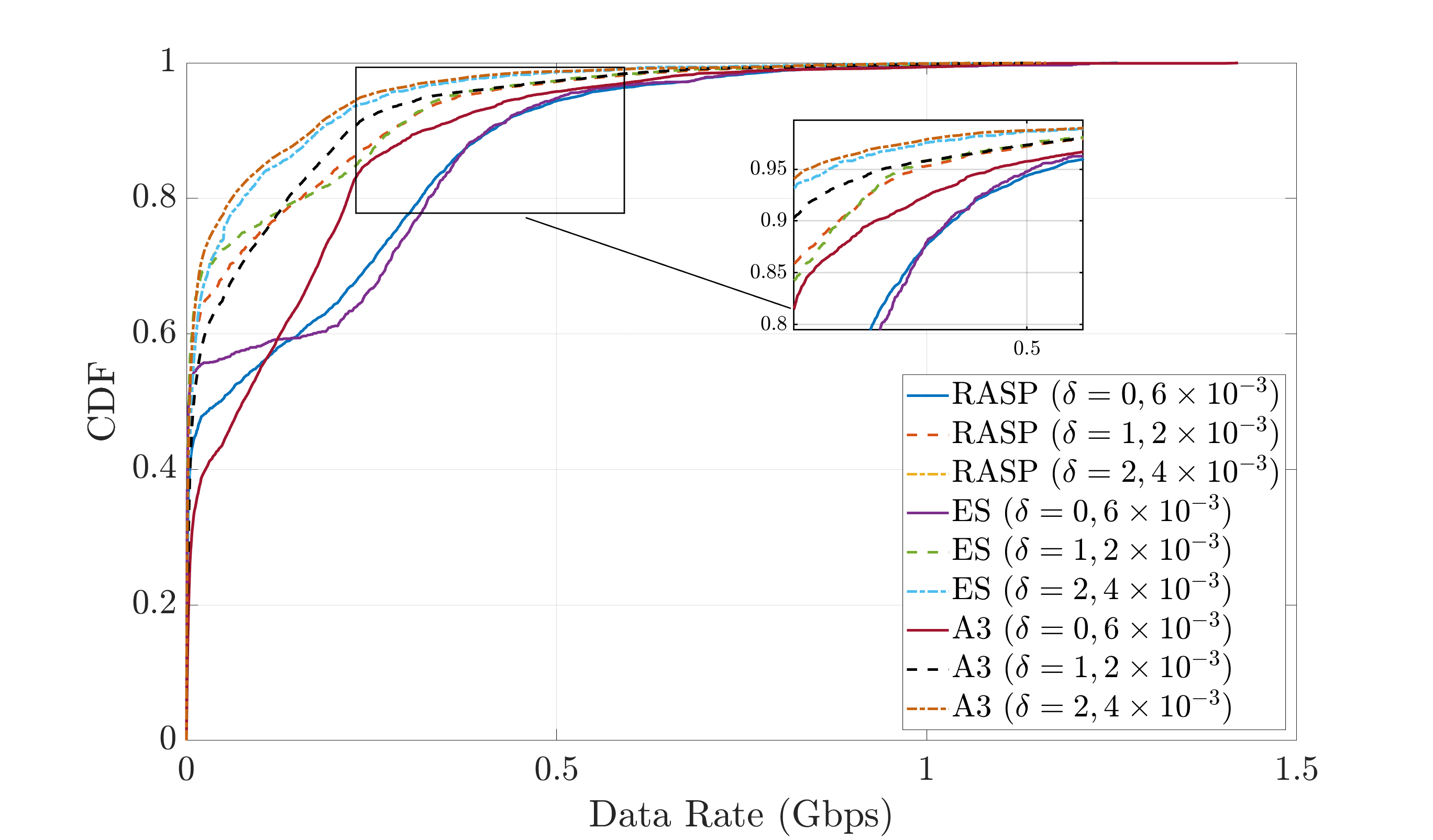}
    \caption{Scenario C when $D = 30\%$.}
    \label{rate_C}
    \end{subfigure}
\caption{Data rate figures for all scenarios and $D = 30\%$ case.}
    \label{Data_Rate}
\end{figure}

Fig. \ref{Data_Rate} demonstrates the data rate results for all scenarios. 
In each scenario, the $D=30\%$ case is considered, because although there is not much difference between 30\% and 70\% in terms of the results, we only present the results of $D=30\%$  for the sake of visualization. 
Fig.~\ref{rate_A} demonstrates the results for Scenario A for two different $D$ values.
In this scenario, since MBS is noise-limited, the UEs hosted by MBS received higher rates than those hosted by other BSs. 
This explains the increase from 500 Mbps to 1.5 Gbps, which classifies the UEs as low-class and high-class in terms of data rate. 
As it can be seen in Figure~\ref{rate_A}, this classification differs as the UE number changes.
Since the number of UEs that MBS can accommodate is limited, as the number of UEs increases, the number of high-class users remains constant, while the number of low-class users increases.
On the other hand, Figs.~\ref{rate_B} and \ref{rate_C} show the results for Scenarios B and C, respectively.
Unlike Scenario A, in these two scenarios, since all BSs are interference-limited, the sharp differentiation between the low-class and high-class is not visible.
The slight rate separation between each number of UEs is also seen in Fig.~\ref{rate_B} and Fig. \ref{rate_C}.

\section{Conclusion}
In this study, a delay-aware CS algorithm for the concept of NTN-integrated CS in 6G networks was proposed for the first time.
HAPS-SMBS and MBS were included in the system model to be utilized in switching off SBSs by accommodating their traffics.
Different network conditions making the communication channels interference and noise limited were considered, and the impacts of two different delay profiles were investigated.
The delay-aware network was tested for different user densities to analyze both sparse and dense network conditions.
Moreover, a novel, lightweight $Q$-learning algorithm design was introduced to present the potential of such a simple algorithm in solving the developed complex CS problem.

\section*{Acknowledgement}
Metin Ozturk was supported by The Scientific and Technological Research Council of Türkiye (TUBITAK) under the TUBITAK-2219 program.

\bibliographystyle{IEEEtran}
\bibliography{output}
\end{document}